# Accurately predicting functional connectivity from diffusion imaging

Cassiano O. Becker[1], Sérgio Pequito[1], George J. Pappas[1], Michael B. Miller[2], Scott T. Grafton[2], Danielle S. Bassett[1,3] and Victor M. Preciado[1†]

**Understanding the relationship between the dynamics of neural processes and the anatomical substrate of the brain is a central question in neuroscience. On the one hand, modern neuroimaging technologies, such as diffusion tensor imaging, can be used to construct *structural graphs* representing the architecture of white matter streamlines linking cortical and subcortical structures. On the other hand, temporal patterns of neural activity can be used to construct *functional graphs* representing temporal correlations between brain regions. Although some studies provide evidence that whole-brain functional connectivity is shaped by the underlying anatomy, the observed relationship between function and structure is weak, and the rules by which anatomy constrains brain dynamics remain elusive. In this article, we introduce a methodology to predict with high accuracy the functional connectivity of a subject at rest from his or her structural graph. Using our methodology, we are able to systematically unveil the role of structural paths in the formation of functional correlations. Furthermore, in our empirical evaluations, we observe that the eigen-modes of the predicted functional connectivity are aligned with activity patterns associated with different cognitive systems. Our work offers the potential to infer properties of brain dynamics in clinical or developmental populations with low tolerance for functional neuroimaging.**

Understanding the relationship between the dynamics of neural processes and the anatomical substrate of the brain is a central question in neuroscientific research [1]. Modern neuroimaging technologies, such as diffusion imaging [2, 3], allow researchers to track white matter streamlines linking cortical and subcortical structures. This information can be conveniently represented in terms of a *structural graph* representing direct anatomical connections [4, 5] between distinct brain regions. Complementary information can be acquired with functional neuroimaging techniques such as functional magnetic resonance imaging (fMRI) [6, 7], which measures time-dependent neural activity in the form of blood-oxygenation-level-dependent (BOLD) signals [8]. Temporal correlations between BOLD signals (averaged over representative brain parcels) can then be used to build a *functional connectivity* matrix, which unveils patterns of global coordination among various brain regions [9]. Prior studies offer preliminary evidence that whole-brain functional connectivity is shaped by the structural graph of anatomical connections [10, 11], yet the extent of this relationship in the human brain is not well understood.

An interesting problem in this context is understanding how functional connectivity emerges from the structural brain graph. This is a challenging problem for several reasons. First, the activity of brain regions that are not directly connected by structural links can be strongly correlated due to indirect structural paths along which signals propagate [12]. Second, the propagation of these signals is influenced, in a nontrivial manner, by the length and number of white matter streamlines in these paths [9]. Furthermore, it is unclear how signals propagating over different structural paths interfere or interact with each other to induce a global pattern of temporal correlations. Several studies have attempted to overcome these difficulties by predicting the functional connectivity of the human brain from the structural graph of anatomical connections. These approaches can be classified in three major groups: (*i*) those performing a direct statistical comparison between the structural graph and the functional connectivity [10, 13, 9, 14]; (*ii*) those based on numerical simulations of brain activity and connectivity [15, 16, 17, 18]; and (*iii*) those using graph-theoretical properties of the structural graph as predictors of functional connectivity [12, 19]. While these studies have made important progress, it remains challenging to accurately predict an individual's functional connectivity from their structural brain graph.

In this article, we introduce a methodology based on spectral graph theory [20] and convex optimization [21] to predict functional connectivity in the resting state (i.e., when a subject is at rest) from the structural graph with high accuracy using a versatile nonlinear predictor. Using our methodology, we are able to systematically unveil the role of indirect structural paths in the generation of functional correlations. In what follows, we describe this predictive methodology and illustrate its performance on neuroimaging data. In our evaluations, we use structural graphs and functional matrices obtained from 84 different subjects measured non-invasively while at rest. In both cases, each node represents an anatomically defined parcel or brain region defined according to the Automated Anatomical Labeling (AAL) atlas [22], which includes 90 cortical and subcortical regions of interest, excluding the brainstem and cerebellar structures. Using diffusion tensor imaging (DTI), we build the edges of the structural graph using the average value of the *Fractional Anisotropy* (FA) [3] over the white matter streamlines con-necting brain regions. The topology of the structural graph can be conveniently represented as an adjacency matrix [23], denoted by $S$, where rows and columns are indexed by brain parcels and the entries are the average FA between each pair of brain parcels. On the other hand, the functional connectivity matrix is computed using functional magnetic resonance imaging (fMRI) of blood-oxygenation-level-dependent (BOLD) time signals [8]. For each brain region, we extract a representative time series using the scale 2 wavelet coefficients (0.06–0.125 Hz) of the mean BOLD signal [9]. An entry $f_{ij}$ in the functional connectivity matrix $F$ is the Pearson's correlation coefficient between representative wavelet coefficients extracted from regions $i$ and $j$.

---
[1] Department of Electrical and Systems Engineering, University of Pennsylvania;
[2] Department of Psychological and Brain Sciences, University of California at Santa Barbara;
[3] Department of Bioengineering, University of Pennsylvania.
† to whom correspondence should be addressed (preciado@seas.upenn.edu)

Using tools from spectral graph theory [20], we propose a technique to predict the functional connectivity matrix $F$ of a subject at rest from his/her structural adjacency matrix $S$. Our predictor is comprised of two stages. In the first stage, we compute a weighted combination of the powers of the structural adjacency matrix $S$ (see **Fig. 1**). As we discuss in the **Online Methods**, the $l$-th power of $S$ accounts for structural paths of length $l$ connecting different brain regions. In practice, we truncate this weighted sum of powers at a particular value $k$, which represents the maximum length of the structural paths taken into account in the functional predictor. In the second stage of our prediction process, we perform a change of coordinates aiming to align the eigenmodes of the structural matrix $S$ with the eigen-modes of the functional connectivity matrix $F$. In algebraic terms, this change of coordinates is performed using a rotation matrix $R$ that depends on the eigenvectors of $S$ and the eigenvectors of $F$ (see **Fig. 1**, and **Online Methods**). As a result of this eigen-mode alignment, we obtain a functional predictor $\hat{F}$ whose entries are a nonlinear combination of measurements related to structural paths of lengths up to $k$.

We compute the parameters of this two-stage functional predictor by solving an optimization problem [21] aiming to maximize the quality of the functional prediction $\hat{F}$. In our experimental evaluations, we measure the functional prediction quality using the Pearson correlation between the entries of the predicted functional matrix $\hat{F}$ and those of the actual functional matrix $F$ (see the **Online Methods**). Hereafter, we refer to this optimization problem as the *spectral mapping problem*. In what follows, we study two different types of spectral mapping problems (represented in **Fig. 2**). First, we consider the problem of finding a 'personalized' functional predictor for each subject. We refer to this problem as the *individual spectral mapping problem*. Second, we consider the problem of finding a 'universal' predictor able to infer the functional connectivity for a whole group of subjects. We refer to this problem as the *group spectral mapping problem*. In what follows, we solve both spectral problems and illustrate the performance of our approach on neuroimaging data acquired from a large cohort of healthy human participants.

## RESULTS

**Individual functional connectivity is predicted with high accuracy**. We first focus our attention on the individual spectral mapping problem (pictorially represented in **Fig. 2** a-d, and mathematically described in the **Online Methods**). In numerical evaluations, we consider two different BOLD time series for each subject, which we construct as follows. Beginning with a single BOLD time series with 146 time samples, we build one set of samples by randomly selecting half the samples from the original signal. The second set is built by choosing the other half of remaining samples. The first set is then used to generate an *in-sample* functional connectivity matrix to train the functional predictor (following the methodology described in the **Online Methods**). The second set is used to generate an *out-of-sample* functional connectivity matrix to validate the quality of the trained predictor. In a first set of experiments, we train and validate 'personalized' functional predictors for each one of the 84 subjects in the dataset. For each individual, we build a hierarchy of predictors with different values of $k$, where $k$ ranges from 1 to 7. In other words, we gradually increase the maximum length of the structural paths being considered in the functional predictor. In **Fig. 3** a, we plot the average quality of the personalized functional predictor for both the training (in-sample) and the validation (out-of-sample) functional matrices. Using the in-sample data, the correlation level achieved by the functional predictor after training becomes consistently close to 1 as $k$ increases above 5 (red boxes in **Fig. 3** a). Using the out-of-sample data to validate the trained predictor, we observe that the quality of the prediction consistently increases with $k$, saturating at an average correlation of 0.79 for $k$ above 5. Based on these results, we can quantify the role of structural paths of different lengths in the formation of the functional connectivity pattern. For example, since the average predictor quality for $k=1$ is 0.562 in the validation dataset, we conclude that direct structural paths of length 1 account for 56.2% of the personalized prediction quality, on average. Furthermore, the average prediction quality increases to 0.686 when we also consider paths of length $k = 2$. Hence, we conclude that structural paths of length 2 account, on average, for a 12.4% increment in the prediction quality (i.e., $100 \times (0.686 - 0.562) = 12.4$). Similarly, as we gradually include structural paths of length 3 to 7 in the predictor, the quality increases according to the following incremental percentages: 3.7%, 4.7%, 1.9%, 0.27%, and 0.12%. Notice how the propagation of neural signals through short structural paths has the strongest influence in the resulting functional connectivity. We also observe that the average prediction quality saturates at $k \geq 5$. Therefore, paths of length up to 5 in the structural graph contain most of the information needed to predict functional correlations, offering fundamental insight into the lengths of paths used for neural computations. In the inset in **Fig. 3 a**, we include a scatter plot to compare the entries of the actual functional connectivity matrix $F_j$ (values in the abscissae) with those of the predicted functional matrix $\hat{F}_j$ (values in the ordinates) when we choose $k=5$ for the individual with the median correlation quality. The ordinates of the red (respectively, blue) dots correspond to the entries of the functional connectivity matrix used for training (respectively, validation).

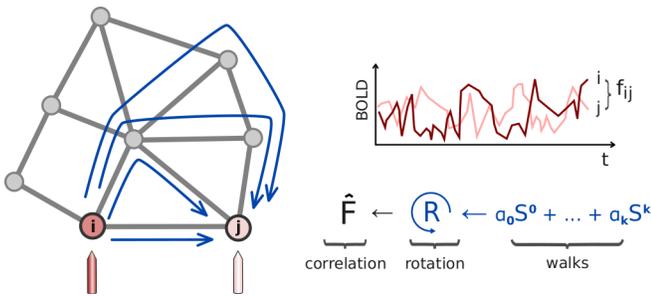

**Figure 1** Approach schematic. The entry $(i,j)$-th entry of the functional connectivity matrix $F$, denoted by $f_{ij}$, represents the correlation between the BOLD time signal wavelet coefficients measured from two brain parcels corresponding to nodes $i$ and $j$ in the functional graph. The $(i,j)$-th entry of the $l$-th power of the structural matrix $S$, denoted by $[S^l]_{ij}$, accounts for walks of length $l$ connecting nodes $i$ and $j$ in the structural graph. In the figure above, several paths of different lengths connecting nodes $i$ and $j$ are indicated by blue arrows. We propose to reconstruct the functional connectivity matrix $F$ using a weighted sum of powers of the structural matrix, denoted by $a_0 S^0 + \ldots + a_k S^k$, as well as a change of coordinates (described by the rotation matrix $R$) aiming to align the eigen-modes of the predicted functional matrix $\hat{F}$ with those of $F$.

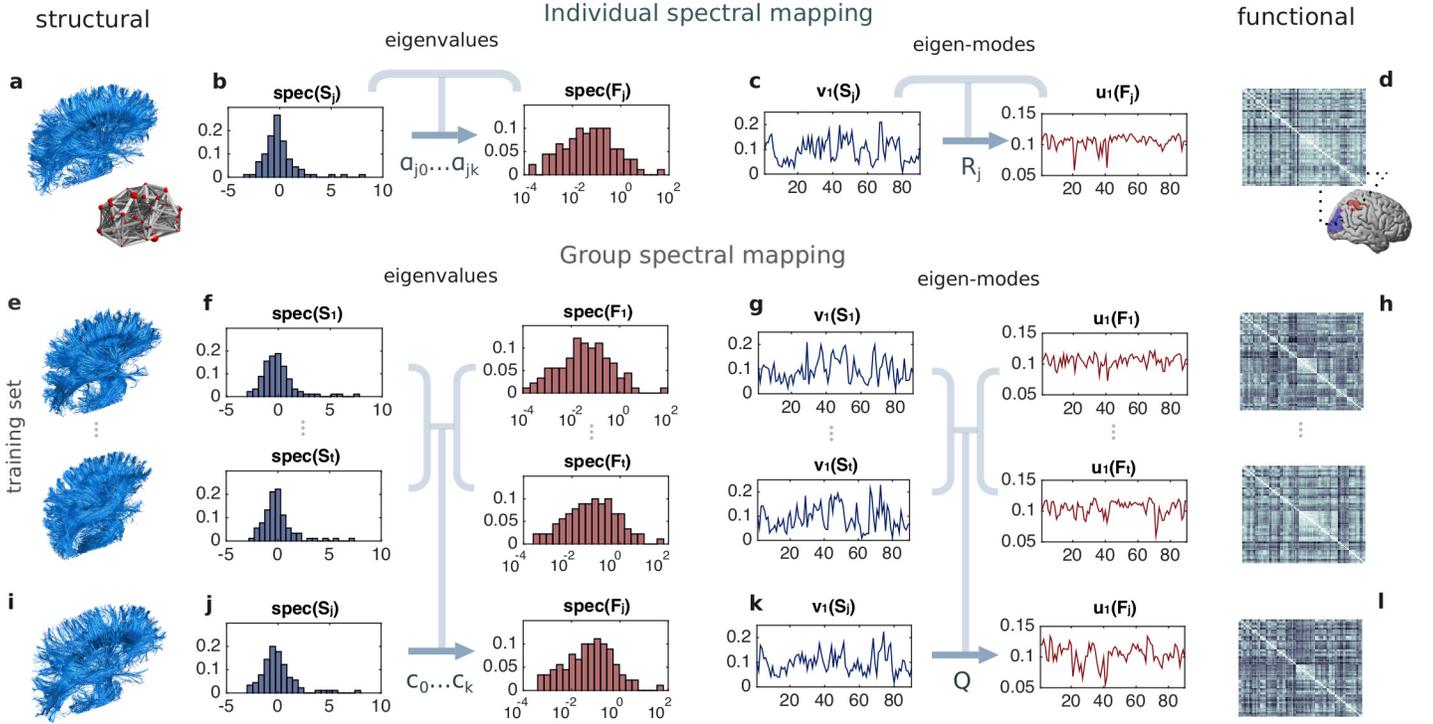

**Figure 2** Spectral mapping method. In the *individual spectral mapping* problem (**a-d**), we predict the functional connectivity matrix $F_j$ of an individual $j$ at rest (**d**) from his/her structural connectivity (**a**). Our prediction is based on a two-stage process. In the first stage, we use a polynomial transformation of order $k$ (characterized by the coefficients $a_{j0},\ldots,a_{jk}$) to predict the eigenvalues of $F_j$ from those of $S_j$. In **b**, we include the histogram of the eigenvalues of $S_j$ and $F_j$ for the $j$-th individual (in linear and log-linear scale, respectively). In the second stage, we use a rotation matrix $R_j$, specific to individual $j$, to infer the eigen-modes of $F_j$ from those of $S_j$ (as illustrated in **c**, for the first eigen-mode). In the lower part of the figure (labels **e-l**), we illustrate the *group spectral mapping* problem, in which we find a 'universal' predictor, valid for a whole group of individuals. For that purpose, we specify a training set composed by structural connectivity graphs (**e**) and their corresponding functional connectivity matrices (**h**). In the first stage (**f**), we find a common polynomial transformation (characterized by $c_0,\ldots,c_k$), using the eigenvalues of structural and functional connectivity matrices in the training set. In the second stage (**g**), we find a common rotation matrix $Q$ aiming to simultaneously align the structural eigen-modes with those of the functional matrices, for all the individuals in the training set. Finally, using both the polynomial transformation (**j**) and the rotation (**k**), we estimate the functional connectivity matrix $F_j$ of a subject $j$ (**l**) from his/her structural graph (**i**).

**Group spectral mapping partially predicts individual connectivity from common parameters**. In addition to the individual spectral mapping problem, we also consider the *group spectral mapping problem*, in which we aim to find a 'universal' predictor able to estimate a representative functional connectivity matrix for a group of subjects. In other words, given the structural graphs and functional connectivity matrices of a group of individuals, we aim to find a *common* predictor to estimate the functional connectivity matrix for any individual in the group with the maximum possible overall correlation quality (see the **Online Methods** for a technical description of this problem). From a neuroscientific perspective, this problem corresponds to understanding the common role of structural paths of differing lengths in neural computations performed in many individuals. To evaluate the performance of the method, we partition the set of 84 subjects into two subsets: an in-sample subset of 42 individuals whose structural graphs and functional connectivity matrices are used to train the universal predictor, and an out-of-sample subset of 42 individuals used to evaluate the prediction quality. In this setup, we find the optimal parameters of the universal predictor for the in-sample training set when the maximum length of structural paths under consideration ranges from $k = 1$ to 7. In **Fig. 3 b**, we plot the performance of the universal functional predictor for both the in-sample training set (yellow boxes) and the out-of-sample validation set (green boxes). Overall, the performance in the validation set is well aligned with the training set, and both tend to increase as we increase the value of $k$. As expected, the universal predictor presents a lower performance when compared with the personalized predictors, highlighting the existence of meaningful individual differences in how structural paths inform functional dynamics. In particular, when the out-of-sample prediction case is considered, the best average performance stabilizes at around 0.62 for the group case (while in the personalized case, it saturates at around 0.79). It is interesting to speculate that differences between the predicted and actual functional connectivity matrices may be related to individual differences in cognitive abilities in healthy individuals, or to individual differences in symptomatology in clinical populations.

**Spectral mapping exhibits robustness to the number of nodes in the parcellation**. It is important to understand whether or not the mappings from structure to function are invariant across spatial resolutions of brain dynamics. To address this question, we performed the same analyses described above using a different anatomical atlas. More precisely, we considered an upsampled version [9] of the Automated Anatomical Labeling Atlas, which we refer to as the AAL-600, developed to create equally-sized regions that still obey gross anatomical boundaries [5, 4]. This upsampled version contains 600 regions created via a series of steps in which regions are bisected perpendicular to their principal spatial axis. Following this process, the resulting atlas contains regions comprised of approximately 268 voxels each. In **Fig. 3 c-d** we present the results for the individual and group spectral mappings in the AAL-600 atlas. Observe that the results are similar to those obtained for the AAL-90 atlas, as illustrated in **Fig. 3 a-b**. In particular, we notice that the performance of the predictor increases as longer structural paths are considered and its performance saturates for structural paths longer than 5.

**Functional eigen-modes are revealed by group spectral mapping**. To investigate the neurophysiological drivers of these predictions, we depict in **Fig. 4** brain surface activation maps representing the first four eigen-modes of the predicted functional connectivity $\hat{F}$ (i.e., the eigenvectors of $\hat{F}$ associated with the four largest eigenvalues). We observe that these eigen-modes are aligned with activity patterns in distinct cognitive systems. In particular, the first eigen-mode (**Fig. 4 a**) represents the so-called Bonacich centrality of the functional connectivity matrix, which measures how 'well-connected' or 'central' a region is in the functional graph [24].

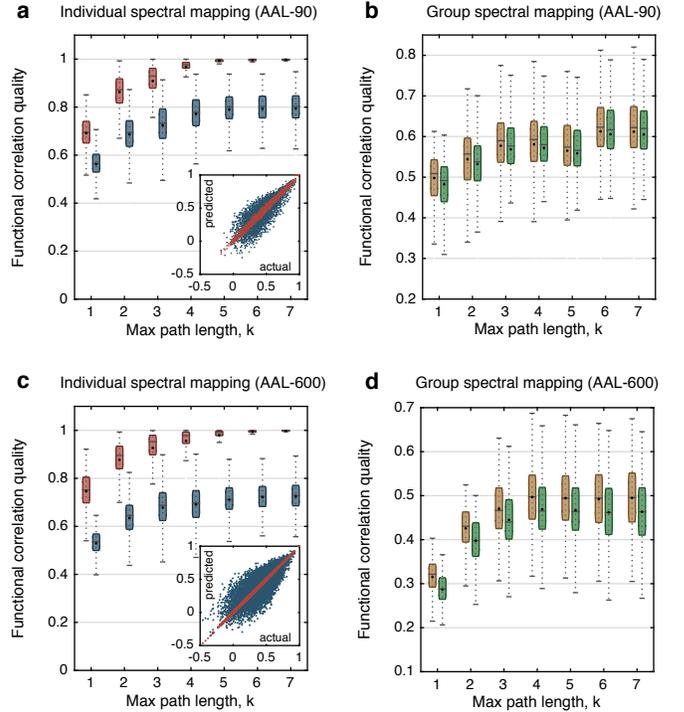

**Figure 3** Spectral mapping performance. We represent the evolution of the correlation quality between the predicted and the actual functional connectivity matrices when we vary the maximum length of the paths under consideration (denoted by the parameter $k$) for the individual and the group spectral mappings. These cases are evaluated for the AAL-90 parcellation in **a**-**b**, and for the AAL-600 parcellation in **c-d**. In **a** (respectively **c**), we plot the evolution of the correlation quality evaluated over 10 different splits of the BOLD signal time-series wavelet coefficients in the training (in red) and validation (in blue) sets. The inset plot in **a** (respectively **c**) includes two scatter plots of the entries of the predicted functional matrix $\hat{F}$ (values in the ordinates) versus the actual connectivity matrix $F$ (values in the abscissae) when $k = 5$ for the individual with the median correlation quality. The ordinates of the red (respectively, blue) dots in the scatter plot correspond to the entries of the functional connectivity matrix used for training (respectively, validation). In **b** and **d**, we plot the evolution of the correlation quality evaluated over 10 different splits of the BOLD signal time-series in the training (in yellow) and validation (in green) sets.

The second eigen-mode (**Fig. 4 b**) takes high values (depicted in yellow) over primary sensory and motor cortices, as well as the adjacent premotor areas and inferior parietal lobule, classically associated with sensorimotor control processes. The third eigen-mode (**Fig. 4 c**) presents high values over primary and secondary visual cortex, as well as the posterior parietal cortex, which are regions associated with visually and somatosensory guided action. Finally, the fourth eigen-mode (**Fig. 4 d**) presents high values over the prefrontal cortex and temporoparietal junction, which are regions associated with high level cognition, attention and the control of behavior.

**Similarity of spectral characteristics across individuals**. From a mathematical perspective, the effectiveness of the group spectral mapping method is, in part, explained by the similarity of spectral characteristics across subjects in the dataset. In particular, as we illustrate below, the eigenvalues of the functional (respectively, structural) matrix present a high level of similarity across individuals. Furthermore, the most relevant eigenvectors of these matrices (i.e., those associated with the largest eigenvalues) are also well aligned among individuals. To evaluate spectral similarities in the data, we start by plotting the eigenvalues of the structural matrix (in decreasing order) in **Fig. 5 c**. In this figure, we include a box plot for the eigenvalues in the dataset. More precisely, for each eigenvalue number (i.e., the $i$-th eigenvalue number of each matrix, in decreasing order), we represent the average value, as well as the first and third quartile for the 84 values corresponding to the $i$-th eigenvalues of all the individuals in the dataset. We observe how the distribution of eigenvalues is very concentrated; in other words, the eigenvalues of the structural matrices in our dataset are very similar across individuals. Similarly, in **Fig. 5 f**, we include a box plot for the eigenvalues of the functional matrices. We observe that, in this case, the eigenvalues are also similar across the dataset. We now shift our attention to the similarity between eigenvectors. In particular, we study the alignment among the first eigenvectors of the structural matrices (i.e., those eigenvectors associated to the largest eigenvalue) across individuals in the dataset. For each pair of individuals, we compute the correlation between their first eigenvectors. Since our dataset contains $l = 84$ individuals, we have $l(l-1)/2 = 3\,486$ possible pairs. In **Fig. 5 a**, we plot a histogram for the values of all these correlations and observe that, in average, the first eigenvectors present a 75% correlation. We repeat this computation using the second eigenvectors of the structural matrices, and plot our results in **Fig. 5 b**. Similarly, in **Fig. 5 d** and **e**, we plot the histogram of correlations for the first and second eigenvectors of the functional matrices, respectively. From a mathematical perspective, these spectral similarities allow us to find a high-quality 'universal' predictor able to map the adjacency matrix of structural brain graphs into functional connectivity matrices using tools from spectral graph theory. From a neurophysiological perspective, such similarities indicate that healthy human subjects display a similar organization of paths in structural graphs, as well as functional connectivity matrices.

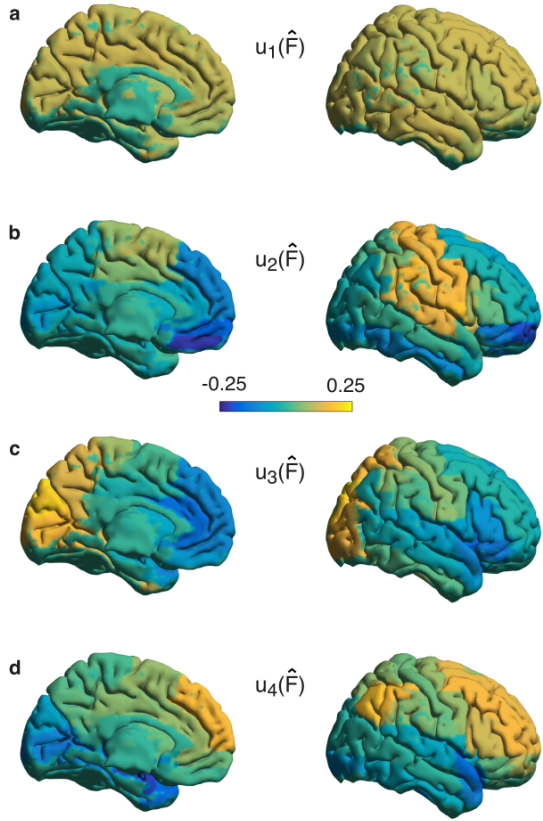

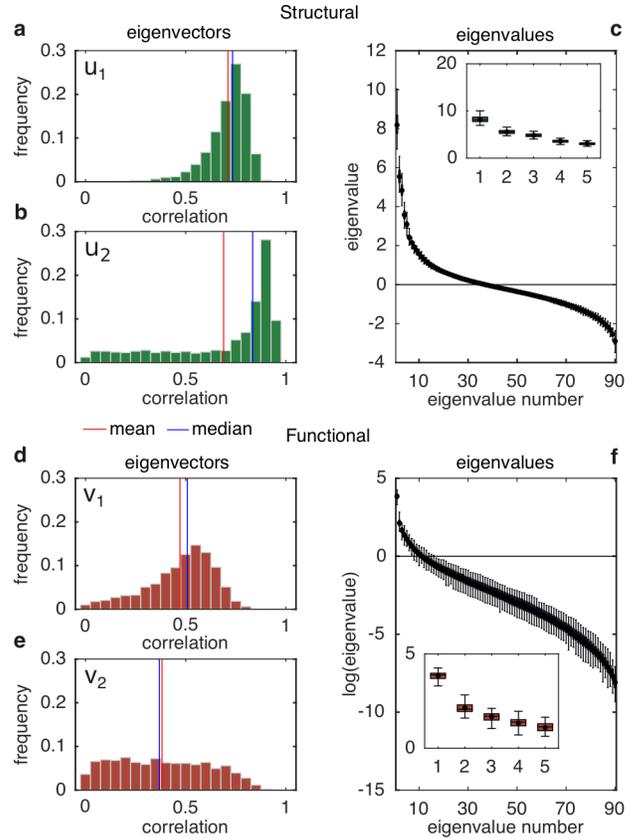

**Figure 4** Eigen-modes of the predicted functional connectivity. Lateral and medial cortical surface renderings [25] for the four eigen-modes associated with the four largest eigenvalues of the functional connectivity matrix predicted by the group spectral mapping.

**Figure 5** Spectral characteristics for structural and functional connectivity matrices. In **a-b** (respectively **d-e**), histograms of eigenvector correlations are displayed for the eigenvectors associated with the two largest eigenvalues of the structural (respectively, functional) matrices. Panel **c** (respectively, **f**) displays the distribution of ordered eigenvalues in the set of structural (respectively, functional) matrices, with the five largest eigenvalues magnified in the inset.

**Stability of spectral mapping**. In what follows, we examine the stability of our predictor to ensure that the spectral mapping method is not overfitting the data. In this direction, we vary the number of degrees of freedom in the functional predictor using two hyper-parameters. The first hyper-parameter is the *order* of the maximum power of $S$, denoted by $k$, included in the first stage of our functional predictor (see **Fig. 1** and **Online Methods**). As previously mentioned, this value is equal to the maximum length of the structural paths considered by the functional predictor. The second hyper-parameter is the *rank* $r$ of the rotation matrix $R$ used in the second stage of the functional predictor (see **Fig. 1** and **Online Methods**). By reducing the rank of the rotation matrix, we restrict the number of eigen-modes of $S$ that we aim to align with the eigenmodes of $F$. Computationally, this has the effect of reducing the number of free parameters associated with the rotation matrix $R$. In **Fig. 6**, we plot the influence of these hyper-parameters on the quality of the solution of the individual spectral mapping problem. This quality is measured as the correlation coefficient between the entries of the functional matrix $F$ and the entries of the predicted functional matrix $\hat{F}$. **Fig. 6** contains subplots displaying the evolution of the functional correlation quality for all the 84 individuals in the dataset as we vary the hyper-parameters. In particular, in **Fig. 6 a** (respectively, **b**), we plot the evolution of the quality when the structural matrix is based on the Fractional Anisotropy (FA) as we vary the value of the hyper-parameter $k$ (respectively, $r$). Similarly, in **Fig. 6 c-d**, we plot the evolution of the functional correlation quality when the structural matrix is built according to an alternative method based on the number of streamlines connecting the regions associated to each pair of nodes [9]. We refer to this alternative type of structural connectivity as *Connect*. In our evaluations, we observe that the functional predictor achieves a high correlation quality even for relatively low values of the hyper-parameters, indicating that it is possible to predict the functional connectivity with a relatively low number of degrees of freedom.

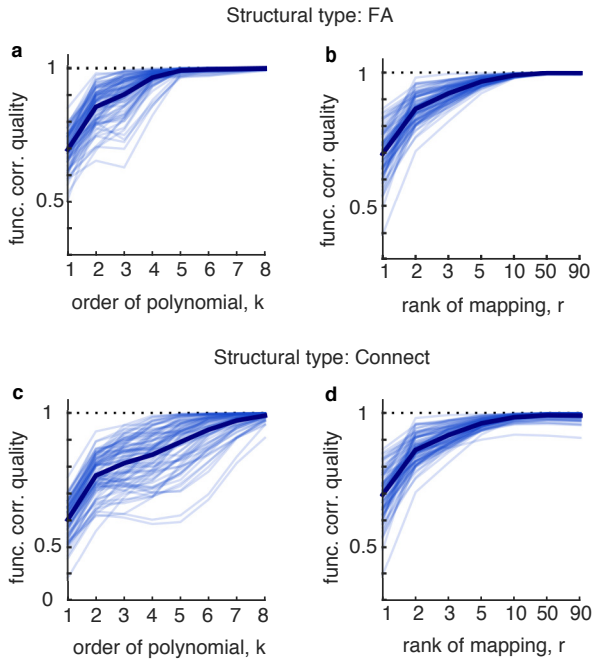

**Figure 6** Stability of spectral mapping. Thin lines denote the evolution of the functional correlation quality for each one of the 84 individuals in the dataset, whereas bold lines indicate the average over all thin lines. In **a-b**, we plot the evolution of the quality for FA-type structural matrices as we vary the value of the hyper-parameters $k$ and $r$. Similarly, **c-d** display this evolution for Connect-type structural matrices.

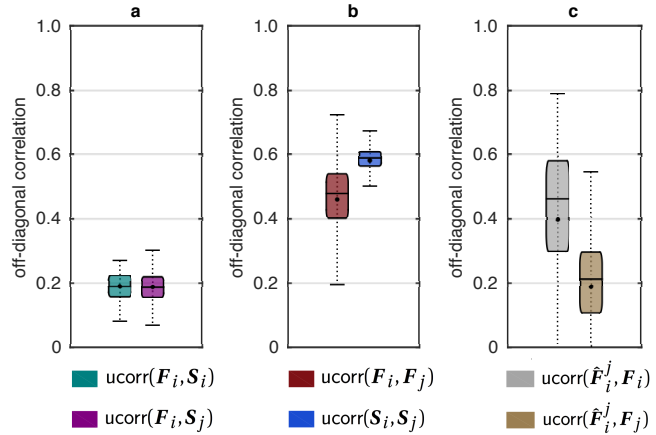

**Figure 7** Null model analysis. This analysis was conducted using structural data of the FA type, while considering full rank spectral mappings and a polynomial of order 8. Briefly, in **a** and **b**, we present the comparison with null models to assess the inherent similarities present in the structural and functional datasets. In addition, in **c** and **d**, we consider the performance of individual spectral mapping and group spectral mapping cases when operating with shuffled data or parameters.

**Comparison to null models**. To gain critical understanding on the effectiveness of our method, we perform a series of tests considering several null models. In our evaluations, we use the function $\text{ucorr}(A, B)$, which is defined as the correlation between the upper-triangular entries of two square symmetric matrices $A$ and $B$ of the same dimension. Using this function, we first evaluate $\text{ucorr}(F_i, S_i)$ for each individual $i$ in the dataset. This correlation measures the inherent similarities between structural and functional modalities for the same subject. In **Fig. 7 a**, we display in green a box plot summarizing the distribution of correlations, which yields a mean (and a standard deviation) of $0.190\,(0.042)$. Apart from similarities between structural and functional matrices for the same subject, we also measure inherent similarities for different subjects. In this direction, we evaluate $\text{ucorr}(F_i, S_j)$ for all pairs $(i, j)$ of individuals in the dataset. In **Fig. 7 a**, we display in purple a box plot summarizing the distribution of correlations, which yields a mean (and a standard deviation) of $0.188\,(0.041)$. These numerical results suggest that the functional matrices $F_i$ in our dataset are poorly correlated with the structural matrices $S_i$; in fact, the average correlation between $F_i$ and $S_i$ is similar to the average correlation between $F_i$ and $S_j$ for $j \neq i$. Furthermore, we also analyze similarities among the functional (respectively, structural) matrices in the dataset by evaluating

ucorr($F_i, F_j$) (respectively, ucorr($S_i, S_j$)) for all pairs $(i, j)$ of individuals in the dataset. In **Fig. 7 b**, we display box plots summarizing the distribution of correlations for the structural and functional modalities, whose means (and standard deviations) are 0.461 (0.112) and 0.582 (0.045), respectively. From these results, we observe that, as expected, structural matrices present a significant level of correlation across individuals, while functional matrices present a lower average correlation.

In the above experiments, we have examined inherent similarities in the dataset by studying correlations between structural and functional matrices for the same, as well as different, individuals. In the following set of experiments, we evaluate the similarity between the 'personalized' functional predictor and functional matrices. As previously described, the 'personalized' functional predictor for the $i$-th individual in the dataset is trained using the pair of matrices ($S_i, F_i$) and characterized by the set of parameters $a_i$ and $R_i$. The input of this predictor is a structural matrix $S$ and its output, denoted by $\hat{F}_i(S)$, is a functional matrix predicted from $S$. In the next experiment, we first compute the functional matrix predicted when the input is the $j$-th structural connectivity $S_j$ using the 'personalized' predictor corresponding to the $i$-th individual. We denote this matrix as $\hat{F}_i^j = \hat{F}_i(S_j)$. We then compare this matrix with $F_i$ and $F_j$ by computing ucorr($\hat{F}_i^j, F_i$) and ucorr($\hat{F}_i^j, F_j$) for all pairs of individuals $(i, j)$ in the dataset. In **Fig. 7 c**, we display in gray a box plot summarizing the distribution of ucorr($\hat{F}_i^j, F_i$), whose mean (and standard deviation) is 0.479 (0.174). This relatively high value is, in part, explained by the inherent similarity among structural matrices of different subjects, as pointed out in the previous paragraph. Similarly, in **Fig. 7 c**, we display in brown a box plot summarizing the distribution of ucorr($\hat{F}_i^j, F_j$) for all $i \neq j$, whose mean (and standard deviation) is 0.226 (0.112). These values should be compared with ucorr($\hat{F}_i^i, F_i$), i.e., the correlation between the predicted and actual functional matrices for the $i$-th individual. As plotted in **Fig. 3 a**, the average correlation in this case is much higher (above 0.99 ($< 0.001$) for the in-sample case, and at 0.79 (0.06) for the out-of-sample case). From these comparisons, it is possible to conclude that the spectral mapping method captures features that are specific to each individual's structural and functional connectivity matrices, since such mapping is not reproducible (on average) by swapping either the matrices or the parameters associated with the mapping.

## DISCUSSION

**Broader implications for cognitive neuroscience.** Our results have important implications for cognitive neuroscience. First, it is striking that paths of length up to $k = 5$ offer maximal prediction accuracy, even across structural brain networks constructed from very different spatial resolutions (from 90 to 600 brain regions). This surprisingly low value of $k$ suggests that relatively parsimonious polysynaptic connections impose critical constraints on brain dynamics and observed functional connectivity. The disproportionate contribution of short paths (of length up to 5) to the functional prediction may be due to energy considerations: it is intuitively plausible that processing information along longer paths may require more energy than processing information along shorter paths. An alternative explanation could lie in the temporal constraints imposed by our environment: over evolutionary time scales, the time it takes an organism to respond to threats or opportunities is negatively correlated with the organism's reproductive success. Assuming longer paths require more time and more energy; it is then reasonable that relatively short paths in the structural graph can predict the observed functional dynamics. However, this line of argument also begs the question of why the functional connectivity matrices cannot be predicted with high accuracy using only paths of length $k = 1$ or $k = 2$. To address this question, it is important to note that paths of increasing length offer a greater dimensionality to the dynamic range of the system. Systems that only utilize structural paths of length $k = 1$ necessarily have an impoverished ensemble of possible states in comparison to those that utilize structural paths of longer lengths. Thus, it is intuitively plausible that the prediction accuracy obtained from $k = 5$ is an indirect indication of a careful balance between the competing requirements for a broad dynamic range and an energetically and temporally efficient system.

**Broader implications for clinical neuroscience.** Our results are also more broadly relevant for clinical neuroscience. Critically, the ability to predict a subject's functional connectivity from their structural connectivity opens the door to inferring brain dynamics even in individuals with low tolerance for functional neuroimaging. Such an ability is particularly relevant in the context of clinical populations with increased motion, anxiety, or proneness to seizures, as well as in developmental populations that have difficulty staying still in the MRI scanner for long periods of time. The principled and accurate inference of brain dynamics from these populations supports the development of personalized therapeutics based on neural markers.

**A generalizable methodological tool.** Finally, it is important to note that the methods we develop here are more generally applicable to other problems in which one wishes to predict one set of matrices from another set of matrices. In the context of neuroimaging, we could use these same tools to ask how structural graphs prior to an injury relate to structural graphs after an injury. We could also ask whether and how functional connectivity matrices change over time, either during learning or as a function of normal aging. It will be interesting in future to determine whether features of the rotation matrix (the mapping from structure to function) are related to individual differences in cognitive abilities in healthy subjects, symptomatology in diseased cohorts, or genetic variability.

**METHODS** Methods and any associated references are available in the online version of the paper.

**ACKNOWLEDGEMENTS** V.M.P. is supported by the NSF under grants CNS-1302222 and IIS-1447470. S.P. is supported in part by the TerraSwarm Research Center, one of six centers supported by the STARnet phase of the Focus Center Research Program (FCRP) a Semiconductor Research Corporation program sponsored by MARCO and DARPA. C.O.B. is supported by CAPES, Coordenação de Aperfeiçoamento de Pessoal de Nível Superior - Brasil. S.T.G. and M.B.M are supported by the Army Research Laboratory through contract no. W911NF-10-2-0022 from the U.S. Army Research Office. D.S.B. acknowledges support from the John D. and Catherine T. MacArthur Foundation, the Alfred P. Sloan Foundation, the Army Research Laboratory and the Army Research Office through contract numbers W911NF-10-2-0022 and W911NF-14-1-0679, the National Institute of Mental Health (2-R01-DC-009209-11), the National Institute of Child Health and Human Development (1R01HD086888-01), the Office of Naval Research, and the National Science Foundation (CRCNS BCS-1441502 and CAREER PHY-1554488).

**AUTHOR CONTRIBUTIONS** S.P., C.O.B. and V.M.P. designed the experiments. S.T.G. and M.B.M. collected the empirical data. Research was conducted by V.M.P. , C.O.B., and S.P., with feedback from D.S.B. and G.J.P.. Data analysis was conducted by C.O.B., S.P., and V.M.P., with feedback from D.S.B., S.T.G. and G.J.P.. The manuscript was written by C.O.B., S.P. and V.M.P., with contributions from D.S.B., S.T.G. and G.J.P..

**COMPETING INTERESTS** The authors declare that they have no competing financial interests.

# References

[1] Sporns, O., Tononi, G. & Kötter, R. The human connectome: a structural description of the human brain. *PLoS Computational Biology* **1**, e42 (2005).

[2] Mori, S. & van Zijl, P. Fiber tracking: principles and strategies–a technical review. *NMR in Biomedicine* **15**, 468–480 (2002).

[3] Gong, G. *et al.* Mapping anatomical connectivity patterns of human cerebral cortex using in vivo diffusion tensor imaging tractography. *Cerebral Cortex* **19**, 524–536 (2009).

[4] Bassett, D. S., Brown, J. A., Deshpande, V., Carlson, J. M. & Grafton, S. T. Conserved and variable architecture of human white matter connectivity. *Neuroimage* **54**, 1262–1279 (2011).

[5] Hagmann, P. *et al.* Mapping the structural core of human cerebral cortex. *PLoS Biology* **6**, e159 (2008).

[6] Huettel, S. A., Song, A. W. & McCarthy, G. *Functional magnetic resonance imaging*, vol. 1 (Sinauer Associates Sunderland, 2004).

[7] Logothetis, N. K. What we can do and what we cannot do with fMRI. *Nature* **453**, 869–878 (2008).

[8] De Luca, M., Beckmann, C., De Stefano, N., Matthews, P. & Smith, S. M. fMRI resting state networks define distinct modes of long-distance interactions in the human brain. *Neuroimage* **29**, 1359–1367 (2006).

[9] Hermundstad, A. M. *et al.* Structural foundations of resting-state and task-based functional connectivity in the human brain. *Proceedings of the National Academy of Sciences* **110**, 6169–6174 (2013).

[10] Greicius, M. D., Supekar, K., Menon, V. & Dougherty, R. F. Resting-state functional connectivity reflects structural connectivity in the default mode network. *Cerebral Cortex* **19**, 72–78 (2009).

[11] Sporns, O. Contributions and challenges for network models in cognitive neuroscience. *Nature Neuroscience* **17**, 652–660 (2014).

[12] Goñi, J. *et al.* Resting-brain functional connectivity predicted by analytic measures of network communication. *Proceedings of the National Academy of Sciences* **111**, 833–838 (2014).

[13] Adachi, Y. *et al.* Functional connectivity between anatomically unconnected areas is shaped by collective network-level effects in the macaque cortex. *Cerebral Cortex* **22**, 1586–92 (2011).

[14] van den Heuvel, M. P., Mandl, R. C., Kahn, R. S., Pol, H. & Hilleke, E. Functionally linked resting-state networks reflect the underlying structural connectivity architecture of the human brain. *Human Brain Mapping* **30**, 3127–3141 (2009).

[15] Honey, C. *et al.* Predicting human resting-state functional connectivity from structural connectivity. *Proceedings of the National Academy of Sciences* **106**, 2035–2040 (2009).

[16] Galán, R. F. On how network architecture determines the dominant patterns of spontaneous neural activity. *PLoS One* **3**, e2148 (2008).

[17] Deco, G., Jirsa, V. K. & McIntosh, A. R. Emerging concepts for the dynamical organization of resting-state activity in the brain. *Nature Reviews Neuroscience* **12**, 43–56 (2011).

[18] Deco, G. *et al.* Resting-state functional connectivity emerges from structurally and dynamically shaped slow linear fluctuations. *The Journal of Neuroscience* **33**, 11239–11252 (2013).

[19] Pernice, V., Staude, B., Cardanobile, S. & Rotter, S. How structure determines correlations in neuronal networks. *PLoS Computational Biology* **7**, e1002059 (2011).

[20] Godsil, C. & Royle, G. F. *Algebraic graph theory*, vol. 207 (Springer Science & Business Media, 2013).

[21] Boyd, S. & Vandenberghe, L. *Convex optimization* (Cambridge University Press, 2004).

[22] Tzourio-Mazoyer, N. *et al.* Automated anatomical labeling of activations in SPM using a macroscopic anatomical parcellation of the MNI MRI single-subject brain. *Neuroimage* **15**, 273–289 (2002).

[23] Newman, M. *Networks: an introduction* (Oxford University Press, 2010).

[24] Bonacich, P. Power and centrality: A family of measures. *American Journal of Sociology* 1170–1182 (1987).

[25] Xia, M., Wang, J., He, Y. *et al.* Brainnet viewer: a network visualization tool for human brain connectomics. *PloS One* **8**, e68910 (2013).

[26] Percival, D. B. & Walden, A. T. *Wavelet methods for time series analysis*, vol. 4 (Cambridge university press, 2006).

[27] Fulton, W. & Harris, J. *Representation theory*, vol. 129 (Springer Science & Business Media, 1991).

## ONLINE METHODS

**Connectivity Matrices.** Structural brain graphs were estimated via deterministic tractography applied on diffusion tensor imaging (DTI) scans [3]. Each region in the structural graph corresponds to a localized brain area in the Automated Anatomical Labeling (AAL) atlas [22]. In our experiments, we consider structural graphs with weighted edges using two types of connectivity: *Connect* and *Fractional Anisotropy* (FA). In the Connect type, the weight of an edge is given by the total number of white matter fiber tracts connecting a pair of brain regions. In the FA type, edge weights are computed using the *fractional anisotropy* of all voxels in the DTI scan, which is a scalar value describing the degree of anisotropy of the diffusion of water molecules in the voxel. The weight of an edge connecting two regions is, thereby, defined as the average value of the fractional anisotropy over the white matter streamlines connecting them [3]. Apart from structural graphs, we also consider functional connectivity matrices, built as follows. For each brain region in the AAL atlas, we extract a representative time-series using the scale-2 wavelet coefficients (0.06–0.125 Hz) of the mean BOLD signal in each region [4]. The $(i,j)$-th entry of the functional connectivity matrix is given by the Pearson's correlation coefficients between the representative time-series of regions $i$ and $j$. Notice that the diagonal entries of the functional connectivity matrices are always equal to one. In our evaluations, we use structural and functional connectivity matrices obtained from 84 different subjects measured non-invasively while at rest.

**Spectral Graph Theory.** Consider a structural brain graph with $n$ nodes and weighted edges. We denote by $S$ the $n \times n$ adjacency matrix of the structural graph, i.e., the entry $[S]_{ij}$ is the weight of the edge connecting nodes $i$ and $j$. We define a path of length $l$ from node $i_0$ to $i_l$ in the graph as an ordered sequence of $l+1$ nodes, $(i_0, i_1, \ldots, i_l)$, such that the pair of nodes $\{i_{r-1}, i_r\}$ are connected for all $r = 1, \ldots, l$. We denote by $\mathcal{P}^l_{i_0, i_l}$ the set of all paths of length $l$ from $i_0$ to $i_l$ in the structural graph. Given a path $p = (i_0, i_1, \ldots, i_l)$, we define the weight of the path as the following product:

$$\omega(p) = [S]_{i_0 i_1} [S]_{i_1 i_2} \ldots [S]_{i_{l-1} i_l}.$$

A fundamental result in spectral graph theory relates the $l$-th power of the adjacency matrix $S$ with paths of length $l$ in the structural graph, as follows:

$$[S^l]_{ij} = \sum_{p \in \mathcal{P}^l_{i,j}} \omega(p). \quad (1)$$

In other words, the $(i,j)$-th entry of $S^l$ is equal to the sum of the weights of all paths of length $l$ from $i$ to $j$. In our solution to the spectral mapping problem, we use a weighted sum of powers of $S$ up to order $k$; in particular, we use the matrix $W = f(S) = a_0 S^0 + \ldots + a_k S^k$. According to (1), the $(i,j)$-th entry of $W$ is equal to a weighted sum over all paths from $i$ to $j$ of length up to $k$. Furthermore, our functional predictor $\hat{F}$ is a similarity transformation of $W$; in particular, $\hat{F} = R W R^\intercal$ with $R$ being a rotation matrix. Therefore, using the spectral mapping theorem [20], we have that the eigenvalues of $\hat{F}$ are given by $a_0 \lambda_i^0 + \ldots + a_k \lambda_i^k$ for $i = 1, \ldots, n$, where $\lambda_i$ is the $i$-th eigenvalue of $S$.

**Individual Spectral Mapping Problem.** In this work, we propose a technique to predict the functional connectivity matrix $F$ of a subject at rest from his/her structural adjacency matrix $S$. As mentioned above, we propose a predictor parameterized as follows: $\hat{F} = R f(S) R^\intercal$, where $f(S) = \sum_{i=0}^k a_i S^i$ and $R$ is an orthogonal rotation matrix. To find the values of the parameters $\{a_i\}_{i=0}^k$ and $R$, we propose to maximize the quadratic (or Frobenius) norm of the difference between $F$ and $\hat{F}$, defined as:

$$\left\| \hat{F} - F \right\|_\mathcal{F}^2 = \sum_{i=1}^n \sum_{j=1}^n \left( [\hat{F}]_{ij} - [F]_{ij} \right)^2.$$

In other words, given the $n \times n$ functional and structural matrices of an individual, $F$ and $S$, as well as the value of the maximum order $k$, we solve the following optimization problem:

$$\begin{aligned}
\underset{\{a_i\}_{i=0}^k, R}{\text{minimize}} \quad & \left\| R \left( \sum_{i=0}^k a_i S^i \right) R^\intercal - F \right\|_\mathcal{F}^2 \quad (2) \\
\text{subject to} \quad & R R^\intercal = R^\intercal R = I_n,
\end{aligned}$$

where the constraints guarantee $R$ to be a rotation matrix. We show in the **Supplementary Information** that the solution to this optimization problem can be found as follows. First, compute the eigenvalues and eigenvectors of $S$ (respectively, $F$), denoted by $\{v_i\}_{i=1}^n$ and $\{\lambda_i\}_{i=1}^n$ (respectively, $\{u_i\}_{i=1}^n$ and $\{\varphi_i\}_{i=1}^n$). Then, mount these eigenvalues into the vectors $\varphi = (\varphi_1, \ldots, \varphi_n)^\intercal$ and $\lambda = (\lambda_1, \ldots, \lambda_n)^\intercal$, as well as the eigenvectors into the matrices $V = [v_1 | v_2 | \ldots | v_n]$ and $U = [u_1 | u_2 | \ldots | u_n]$. Define the Vandermonde matrix

$$L = \begin{bmatrix} 1 & \lambda_1 & \lambda_1^2 & \cdots & \lambda_1^k \\ 1 & \lambda_2 & \lambda_2^2 & \cdots & \lambda_2^k \\ \vdots & \vdots & \vdots & \ddots & \vdots \\ 1 & \lambda_n & \lambda_n^2 & \cdots & \lambda_n^k \end{bmatrix},$$

where the parameter $k$ is the maximum order of the polynomial. Therefore, it can be shown (see **Supplementary Information**) that the solution pair $(\{a_i^*\}_{i=0}^k, R^*)$ to the optimization problem in (2) is given by $(a_0^*, \ldots, a_k^*)^\intercal = (L^\intercal L)^{-1} L^\intercal \varphi$ and $R^* = U V^\intercal$. Subsequently, the functional predictor is given by:

$$\hat{F} = R^* \left( \sum_{i=0}^k a_i^* S^i \right) (R^*)^\intercal.$$

In what follows, we explain how to generate in-sample and out-of-sample functional matrices to train our predictor and assess its quality. We start with a collection of bold signals $b_r = (b_r(1), \ldots, b_r(146))^\intercal$ of dimension 146, where $b_r(s)$ denotes the $s$-th time sample of the average BOLD signal in the $r$-th brain region (according to the AAL atlas with either 90 or 600 regions). We then compute the scale-2 maximum-overlap wavelet transforms [26] of $b_r$ for all $r$, which we denote by $w_r = (w_r(1), \ldots, w_r(146))^\intercal$. For each region $r$, we partition the wavelet vector $w_r$ into two: one first vector $w_r^{(1)}$ including 73 entries of $w_r$ (chosen uniformly at random without repetition), and $w_r^{(2)}$, which includes the remaining entries of $w_r$. Using the sets of vectors $\{w_r^{(1)}\}_{r=1}^{90}$ and $\{w_r^{(2)}\}_{r=1}^{90}$, we compute two functional correlation matrices; the *in-sample* matrix $F^{(1)}$ and the *out-of-sample* matrix $F^{(2)}$, where $[F^{(1)}]_{ij}$ (respectively, $[F^{(2)}]_{ij}$) is the Pearson correlation coefficient between $w_i^{(1)}$ and $w_j^{(1)}$ (respectively, $w_i^{(2)}$ and $w_j^{(2)}$). In our numerical experiments, we use $F^{(1)}$ to find the optimal set of parameters for the predictor $\hat{F}$ (i.e., we use $F^{(1)}$ to solve the optimization problem in (2)). In the main document, we use the out-of-sample matrix $F^{(2)}$ to assess the quality of our predictor.

Since an $n \times n$ rotation matrix $R$ has $n(n-1)/2$ degrees of freedom [27], the solution to (2) may result in overfitting due to the additional number of parameters in the predictor. To validate that this is not the case, we consider a version of the optimization problem (2) in which we constrain the rank of $R$ to be equal to a given integer $\rho$ (see **Supplementary Information** for more details). In our numerical experiments, we compute the predictor for different values of $\rho$ and plot our results in **Fig. 6**. Our results support our claim about the absence of overfitting in $\hat{F}$.

**Group Spectral Mapping.** Consider a group of $N$ individuals whose structural and functional matrices are given by the set of pairs $\{(S_j, F_j)\}_{j=1}^N$. In this mapping problem, our objective is to find a *common* predictor able to generate an approximation of $F_j$ from $S_j$. Our predictor is parameterized as follows: $\hat{F}_j = R f(S_j) R^\intercal$, where $f(S_j) = \sum_{r=0}^k a_r S_j^r$ and $R$ is an orthogonal rotation matrix. To find the parameters $\{a_r\}_{r=0}^k$ and $R$ in the common predictor, we propose to solve the following optimization problem: given $N$ pairs of matrices $\{(S_j, F_j)\}_{j=1}^N$, as well as the value of the maximum order $k$, find the solution pair $(\{a_r^*\}_{r=0}^k, R^*)$ that solves

$$\begin{aligned}
\underset{\{a_r\}_{r=0}^k, R}{\text{minimize}} \quad & \sum_{j=1}^N \left\| R \left( \sum_{r=0}^k a_r S_j^r \right) R^\intercal - F_j \right\|_\mathcal{F}^2 \quad (3) \\
\text{subject to} \quad & R R^\intercal = R^\intercal R = I_n.
\end{aligned}$$

This is a hard, non-convex optimization problem and we propose an efficient approximation algorithm in the **Supplementary Information**. Our dataset consists of a group of 84 subjects. In our experiment, we partition this group into two subgroups of equal size. We use the pairs of structural and functional matrices in the first subgroup (which we refer to as the *in-sample* set) to train the common predictor, and validate our results with the second subgroup (the *out-of-sample* set).

**Matrix Correlation Quality.** In our empirical evaluations, we measure the similarity between two $n \times n$ square matrices $X$ and $Y$ using the *matrix correlation function*, denoted by $\text{ucorr}(X, Y)$, and defined as the entry-wise correlation between the upper-triangular entries (excluding diagonal elements) of $X$ and $Y$. In other words, if we build two vectors $x$ and $y$ of dimension $n(n-1)/2$ by 'vectorizing' the upper triangular entries of $X$ and $Y$, the matrix correlation function is simply the correlation between these two vectors.